\documentclass[aps,prl,amsfonts,twoside,twocolumn,amssymb,superscriptaddress,dvips,showpacs]{revtex4}

\usepackage{amsmath}
\usepackage{amsfonts}
\usepackage{amssymb}
\usepackage{graphicx}

\usepackage{bbm}

\def\ket#1{\vert#1\rangle}

\begin{document}

\title{A universally programmable Quantum Cellular Automaton}
\author{D.~J. Shepherd} \affiliation{Department of Computer Science, University of Bristol, Merchant Venturers Building, Bristol BS8 1UB U.K.} 
\author{T. Franz} \affiliation{Institut f\"ur Mathematische Physik, TU Braunschweig, Mendelssohnstra\ss e 3, D-38106 Braunschweig, Germany} 
\author{R.~F. Werner} \affiliation{Institut f\"ur Mathematische Physik, TU Braunschweig, Mendelssohnstra\ss e 3, D-38106 Braunschweig, Germany} 

\begin{abstract}
We discuss the role of classical control in the context of reversible quantum cellular automata. Employing the structure theorem for quantum cellular automata, we give a general construction scheme to turn an arbitrary cellular automaton with external classical control into an autonomous one, thereby proving the computational equivalence of these two models. We use this technique to construct a universally programmable cellular automaton on a one-dimensional lattice with single cell dimension 12.
\end{abstract}

\pacs{03.67.Lx}

\maketitle


Many architectures have been proposed for the construction of a quantum computer. While the earliest algorithms were considered in a model based on unitary gates, recent years have seen ideas like the one-way quantum computer, in which the non-unitary acts of measurement play the key role, or adiabatic computing, in which the continuous time dynamics plays a key role. It is quite possible that substantially new ways quantum computation will surface. 

Studies showing how different quantum computational models can simulate each other are valuable in constructing universal paradigms.  They also provide perhaps the clearest expression of the primitives in each computational model that are responsible for generating computational power stronger than that of classical computation. Since the major obstacles against quantum computation will for a long time be difficulties in implementation, a further incentive for such alternatives is to generate ideas of how to adapt the computational model to an available set of primitives.

In this paper we chiefly consider two different computational models, both of which are quantum cellular automata (QCAs), i.e., distributed systems of lattice cells with a spatially homogeneous discrete time dynamical evolution of strictly finite propagation speed. We use the definition for QCAs developed in \cite{lit:QCA}. This differs from an older definition given in \cite{lit:Watrous}, which is not consistent with the iteration of dynamics (see chapter V in \cite{lit:QCA}).

Our two models differ from each other in the way the program operates, or more precisely how the quantum part of the computer interacts with a classical controller, being somewhat analogous to the gate model and the Turing machine model respectively.
In the gate model the classical controller has to be very powerful: on receiving the input, it will compile the program in a version adapted to the size of the input, and actually build a quantum circuit to run it. The flexibility of this model hence largely resides in the classical controller, and the quantum computer hardware is, in principle, scrapped after each run. 

In contrast, a classical universal Turing machine takes its flexibility from the possibility of writing both the program and the input data on its tape for initialization. We can apply these ideas to running a quantum cellular automaton as a computer: on the one hand, we can use a classical compiler to select a classical sequence of operations each of which is a QCA time step in its own right. Such a machine will be called a {\em classically controlled QCA} (ccQCA). On the other hand, we can insist that program and data are written into the system by the initial preparation, after which the machine runs autonomously for a fixed number of steps, and a fixed transition rule independent of the program. The only role of the classical controller is then final measurement.

It is entirely possible that the absence of classical signalling from this second model (except at initialisation and readout,) coupled with temporal translational symmetry, will prove to have the pragmatic value that an implementation can be more readily isolated from decoherence channels while it is `running' its program, thereby enabling lengthy computation without explicit error-correction.

Our aim is to show that these two ways of programming a QCA are computationally equivalent. In the proof we use the structure Theorem for cellular automata obtained in \cite{lit:QCA}. We then use this equivalence to build a universal autonomous QCA, with an explicitly given transition rule, where ``universal'' means that it simulates the gate model up to polynomial overhead. This construction employs a significantly smaller cell size than that of other similar machines discussed in the literature, \cite{lit:Raussendorf2, lit:Vlasov}. 

The structure Theorem holds in any lattice dimension, and so do the ideas of our construction, but we stick to the one-dimensional case as it is sufficient for bounded error quantum probabilistic computation, ({\bf BQP}). The advantage of using just one lattice dimension has also been stressed by some authors \cite{lit:Benjamin} for practical engineering concerns. However, first implementations of  higher dimensional lattices already exist \cite{lit:Bloch}.

\subsection{General construction techniques} \label{General}

\subsubsection{From gate model to ccQCA}
Consider the gate model of quantum computation wherein qubits are
present in a 1-dimensional lattice, and any gate may act unitarily
on just two neighbouring qubits. Such models are universal in the
sense of \textbf{BQP} computation \cite{lit:QGC}. Then there are
various direct ways of implementing such circuits as classically
controlled QCAs: for example, one could envisage a data band for
encoding the qubits of a circuit, and a pointer band for encoding
a single data-pointer. The transformations of the ccQCA could
manipulate the location of the pointer and then use that pointer
to break the spatial symmetry of the dynamics so that individual
specific neighbouring qubit pairs may be addressed, as required. The data band and
pointer band can of course be regarded as one band, by
interleaving their qubits. 

\subsubsection{From ccQCA to QCA}
We now give a construction, which turns an arbitrary ccQCA with finite
instruction set into a QCA without classical control. The design
of the data band of the ccQCA will be retained, and its programs
will be encoded in an additional band of quantum cells.

\begin{figure}[t]
     \includegraphics[width=7.5cm]{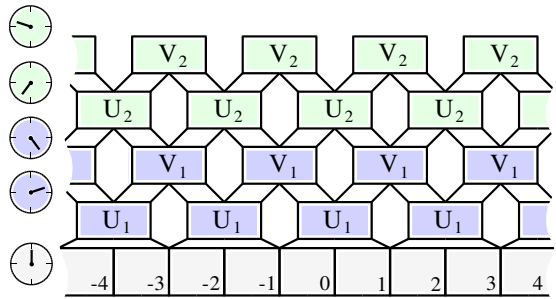}
     \caption{\label{fig:ccQCA} Two timesteps of the ccQCA. The clocks indicate the the time before and after application of the unitaries for comparison to FIG. 2.}
\end{figure}

The main tool needed for the decomposition is the QCA structure
theorem (Theorem 6 in \cite{lit:QCA}). This theorem guarantees the
existence of two unitary operations ($U_i$ and $V_i$) for each of
the ccQCA transition rules which implement the time evolution by
sequential application to non-overlapping neighbourhoods (as
indicated in \mbox{FIG. \ref{fig:ccQCA}}). The structure theorem
applies directly to nearest neighbour ccQCAs. To apply it to a ccQCA
with larger neighbourhood, one needs to group adjacent cells of the
ccQCA into "super-cells" such that one ends up with a
nearest-neighbour interaction for the super-cell structure. Note
that the interaction is not changed by this way of reorganizing the
QCA, and the neighbourhood of the regrouped ccQCA is only slightly
enlarged to become a union of super-cells. In any case, the analysis can be 
restricted to nearest neighbour automata.  

\begin{figure}
     \includegraphics[width=7.5cm]{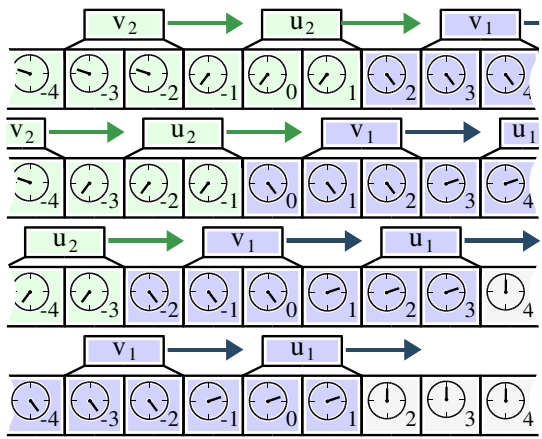}
     \caption{\label{fig:QCA} Four timesteps of the autonomous QCA. As the cells are updated sequentially, the clocks indicate the \textit{local time} of the cells corresponding to the time of the ccQCA.}
\end{figure}

Consider an autonomous QCA consisting of a data band representing the one-dimensional lattice of the \mbox{ccQCA} being simulated and a program band containing information about the sequence of transition functions that the ccQCA would apply.  Let the ccQCA contain $k$ different homomorphisms. Then the cell size of the program band is chosen to be $2k + 1$, enough to distinguish the $2k$ unitaries, allowing for an extra symbol representing the identity map.  Let the time evolution of the QCA be the product of a `shift step' shifting the program band two cells past the data band, followed by a `calculation step' performing the required unitary maps on pairs of cells of the data band, each controlled by the neighbouring contents of the program band.
As the program band moves past the data band, each data cell (qudit) undergoes the time evolution of the \mbox{ccQCA} being simulated, yet it should be noted that different timesteps in the \mbox{ccQCA} evolution are present at one timestep of the autonomous QCA (see \mbox{FIG. \ref{fig:QCA}}).  In accordance with the definition of QCA, it is important that there arises no possibility of non-commuting unitaries operating on the same cell at any time; note that the unitaries $U_i$ and $V_i$ obtained from the QCA structure theorem work on different combinations of odd and even cells (see \mbox{FIG. \ref{fig:ccQCA}}).  Therefore, to circumvent this possibility, one can design the autonomous QCA such that the localisation regions of the unitaries are separated by one \textit{idle} cell, as depicted in our example.

This construction gives an autonomous QCA which, since its dynamics must by definition be (spatially) translationally symmetric, has cells composed of three cells of the original ccQCA plus one cell from the program band; and it turns out to have only nearest-neighbour interactions.  This general construction scheme can be optimized in an explicit situation to reduce the large cell-size.  Next we give such an explicit construction by starting from a universal \mbox{ccQCA} with homomorphisms that already have sequential structure, so the Margolus decomposition can be omitted.

\subsection{Explicit construction} \label{Explicit}
\subsubsection{Construct circuit of controlled partial $\sigma_y$}
There is a two-qubit gate which, if not constrained to act always on neighbouring qubits but allowed to act on qubits within arbitrary range, serves as universal for computation within the standard gate model.  This gate is defined as 
\begin{equation} \label{def:G}
  G ~=~ \left( \begin{array}{cccc}
     1&0&0&0\\0&1&0&0\\0&0&\frac{1}{\sqrt2}&\frac{-1}{\sqrt2}\\0&0&\frac{1}{\sqrt2}&\frac{1}{\sqrt2} \end{array} \right)
\end{equation}
in the computational basis, $\{\,\ket{00}, \ket{01}, \ket{10}, \ket{11}\,\}$. It performs a $\pi/4$ rotation on one qubit conditioned on the setting of another.
To see that this gate is universal, it suffices to show that by use of simple (computational-basis) ancill\ae, one can simulate both the Hadamard and the Toffoli gates, according to the well-known results \mbox{of \cite{lit:QGC}.}

\subsubsection{Construct qubit ccQCA}
\begin{figure}
     \includegraphics[width=5cm]{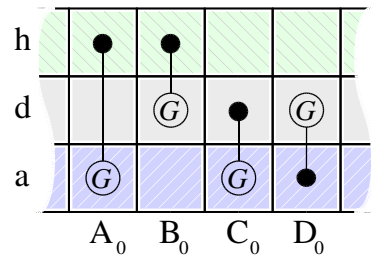}
     \caption{\label{fig:three_bands} Structure of the `three band' QCA with pointer, data and ancilla band. Localisation of the homomorphisms (\ref{homomorphisms1}) are indicated.}
\end{figure}

Consider a ccQCA on a 1-dimensional qubit lattice that allows the use of four different kinds of QCA-homomorphisms as described below.
To show how this ccQCA can be used to simulate an arbitrary gate model circuit whose gates are of the kind $G$ (\ref{def:G}), we will think of the \mbox{ccQCA's} qubits as belonging to three interleaved 1-dimensional lattices, and label them accordingly as $d_i, a_i, h_i$ with $i \in \mathbbm{Z}$.
We will load the $d$-band with input corresponding to the input of the gate array being simulated, initialising its unused qubits to $\ket0$.
The $a$-band will be used as `ancilla space' and should be initialised to $\ket0$ everywhere.
The $h$-band is used to break spatial symmetry of the dynamics, and should contain a single `pointer' $\ket1$ with the rest of its qubits containing $\ket0$.
The four families of homomorphisms we allow are
\begin{eqnarray} \label{homomorphisms1}
  A_i=\prod_{x \in \mathbbm{Z}} G(h_x,a_{x+i}),&&
  B_i=\prod_{x \in \mathbbm{Z}} G(h_x,d_{x+i}), \nonumber\\\\
  C_i=\prod_{x \in \mathbbm{Z}} G(d_x,a_{x+i}),&&
  D_i=\prod_{x \in \mathbbm{Z}} G(a_x,d_{x+i}),\nonumber
\end{eqnarray}
(see FIG.\ref{fig:three_bands}) and we simulate the gate $G( d_i,
d_j )$ by the classically controlled sequence (reading right to
left)
\begin{equation}
A_0^7\cdot C_{-i}^7 \cdot B_i^6 \cdot C_{-i} \cdot D_j \cdot C_{-i}^7 \cdot B_i^2 \cdot C_{-i} \cdot A_0 = G(d_i, d_j).
\end{equation}
The reader may check that this effects the required transformation on the $d$-band, restoring the other two bands to their original configurations, assuming the initialisations described above.  Of course, a clever compiler would find ways of simulating a given circuit that are far more efficient than repeated application of this technique.

\subsubsection{Construct nearest neighbour qubit ccQCA}
The ccQCA described above employs operations with arbitrarily large
neighbourhood. With additional neighbour-swap operations we can
first move two bands to the required interband-distance $i$, then
apply a cellwise $G$-operation ($A_0,B_0,C_0$, or $D_0$) and
finally shift back, in order to implement all operations
(\ref{homomorphisms1}) with a nearest neighbour ccQCA. Moreover, the
interband $G$-operations are quite similar, which suggests to
interleave the three bands into a single qubit band labelled $q_i, i \in \mathbbm{Z}$ 
$$ (..,q_{-2},q_{-1},q_0,q_1,q_2,..)=
   (..,a_{-1},h_{-1},d_0,a_0,h_0,..).$$
A sufficient set of operations is then given by
\begin{eqnarray} \label{homomorphisms2}
  E_i & = & \prod_{x \in \mathbbm{Z}} \mbox{Swap}(q_{3x+i},q_{3x+i+1}), \nonumber \\
  F_i & = & \prod_{x \in \mathbbm{Z}} G(q_{3x+i},q_{3x+i+1}),
\end{eqnarray}
for $i \in \{\,0,1,2\,\}$. It should be noted that it suffices to move two bands relative to each other, since the homomorphisms of (\ref{homomorphisms1}) act non-trivially on only two bands at a time.

\subsubsection{Construct universal nearest-neighbour QCA}
The homomorphisms of the ccQCA already work on
non-overlapping neighbourhoods, and so there is no need for further
Margolus decomposition here. Next we offer a design which further
minimizes the Hilbert space dimension of the cells of which
the QCA will be composed.

\begin{figure}
     \includegraphics[width=7cm]{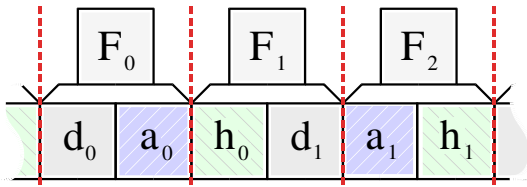}
     \caption{\label{fig:interleaved} Structure of the autonomous QCA.
     Indicated are three cells with one program qutrit and two data qubits each;
     identification of the data qubits corresponding to \mbox{FIG. \ref{fig:three_bands}}.}
\end{figure}

Take a 1-dimensional lattice of qudits labelled $c_i$ (for \mbox{$i \in \mathbbm{Z}$}) of single cell dimension $d=12$, and regard these as incorporating one qutrit cell of the program band $t_i$ with two qubit cells $q_{2i}$ and $q_{2i+1}$ from the data band.  The cell $c_i$ we define explicitly as the tensor product
\begin{equation}
  c_i = t_i \otimes q_{2i} \otimes q_{2i+1}.
\end{equation}
(Identification of data cells is indicated in FIG.
\ref{fig:interleaved}) As before, it is not necessary to have any
`fine control' over the relative motion of the two sub-bands; rather
we simply allow one to pass by the other with an invariant
velocity.  This is again achieved by decomposing the QCA
transformation step into two parts, a unitary and a shift:
\begin{eqnarray} \label{def:finalTransform}
  U\colon~ \mathbbm{C}^{12} \rightarrow \mathbbm{C}^{12} \mbox{ acting on every cell simultaneously,} \nonumber \\
  S\colon~ \left( \begin{array}{rcl}t_i&\mapsto&t_{i+1}\\
  q_i&\mapsto&q_{i-1} \end{array} \right) \mbox{ sliding the bands relatively.~~~~}
\end{eqnarray}

To simulate the nearest-neighbour ccQCA, we will
interpret the data band $q_i$ exactly as before, but the program
band $t_i$ must be initialised so as to execute the appropriate
transformations on the data band as the two bands slide past one
another. At initialisation, the cells $i>0$ will be used to hold
the non-zero content of the data band in their qubits, while the
cells $i<0$ will be used to hold the program band in their
qutrits. We will initialise the $t_i$ in the computational basis,
and the $U$ operation will be defined to leave these qutrits invariant.
Specifically, $t_i=\ket0$ will cause no transformation,
$t_i=\ket1$ will cause a swap of data between $q_{2i}$ and
$q_{2i+1}$, and $t_i=\ket2$ will cause the transformation
$G(q_{2i}, q_{2i+1})$ described \mbox{in (\ref{def:G}).}

Each of the homomorphisms of (\ref{homomorphisms2}) is simulated
not by one QCA timestep but by three neighbouring qutrits sliding
past all of the data-bearing qubits. Specifically, the
program-segment
 $\ket{t_{3i}\,t_{3i+1}\,t_{3i+2}}\equiv\ket{100}$
will simulate the homomorphism $E_0$, the program-segments
$\ket{010}$ and $\ket{001}$ will simulate the homomorphisms $E_2$ and $E_1$.
Likewise, the program-segments $\ket{200}$, $\ket{020}$,
$\ket{002}$, will simulate the homomorphisms $F_0$, $F_2$, $F_1$,
respectively. The cells with negative index may be initialised with
program-segments of these kinds in order to induce the desired
transformations on the data.  The computation output may be read (in computational
basis) any time after the content of the program
band has moved past the content of the data band.

To show that this simulation is efficient, one needs to estimate
the necessary resources. Consider a quantum gate circuit (QGC) of
\textit{Space}$_{QGC}$ qubit-wires and \textit{Time}$_{QGC}$
$G$-gates. In the first simulation step, the resources of the
ccQCA depend linearly on the corresponding QGC resources. The use
of swap gates in the next step increases the time; the encoding of
the program into the program band increases the space, so one ends
up with an estimate for the autonomous QCA of
\begin{eqnarray}
\mbox{\textit{Time}}_{QCA} = \mathcal{O}(\mbox{\textit{Time}}_{QGC}\cdot \mbox{\textit{Space}}_{QGC}), \nonumber \\
\mbox{\textit{Space}}_{QCA} = \mathcal{O}(\mbox{\textit{Time}}_{QGC}\cdot \mbox{\textit{Space}}_{QGC}).
\end{eqnarray}
The resources depend polynomially on the given QGC, so the
simulation is efficient.

\subsection{Acknowledgements} \label{Ack}
DJS is sponsored by CESG. RFW acknowledges support from the QUPRODIS network of the EU.



\begin{thebibliography}{99}
 \bibitem{lit:QCA}
    B. Schumacher and R.F. Werner, \textit{Reversible quantum cellular automata}, (quant-ph/0405174)
 \bibitem{lit:Watrous}
    J. Watrous, \textit{On one-dimensional quantum cellular automata},
    in \textit{Proceedings of the 36th Annual Symposium on Foundations of
    Computer Science} (1995) pp.~528--537 
 \bibitem{lit:Raussendorf2}
     R. Raussendorf, \textit{A quantum cellular automaton for universal quantum computation}, Phys. Rev. A \textbf{72}, 022301 (2005) 
  \bibitem{lit:Vlasov}
 A. Y. Vlasov, \textit{Programmable Quantum Networks with Pure States}, (quant-ph/0503230)
 \bibitem{lit:Benjamin}  S.~C.~Benjamin, S.~Bose,
  \textit{Quantum Computing in Arrays Coupled by `Always On' Interactions}, Phys. Rev. A \textbf{70}, 032314 (2004) 
 \bibitem{lit:Bloch}  I.~Bloch,
  \textit{Ultracold Quantum Gases in Optical Lattices },  Nature Physics \textbf{1} (2005) pp.~23--30
  \bibitem{lit:QGC}
    D. Aharonov, \textit{A simple Proof that Toffoli and Hadamard are Quantum Universal}, (quant-ph/0301040)
\end{thebibliography}
\end{document}